\documentclass[a4paper,12pt]{article}
\usepackage{epsfig}
\usepackage{epstopdf}
\usepackage[sumlimits,intlimits,namelimits]{amsmath} 

\usepackage[english]{babel}

\usepackage{graphicx}

\usepackage{geometry}
\numberwithin{equation}{section}
\newcommand{\beq}{\begin{equation}}
\newcommand{\eeq}{\end{equation}}
\newcommand{\bea}{\begin{eqnarray}}
\newcommand{\eea}{\end{eqnarray}}

\newcommand{\tto}{\!\to\!}

\newcommand{\gsim}{\lower.7ex\hbox{$
\;\stackrel{\textstyle>}{\sim}\;$}}
\newcommand{\lsim}{\lower.7ex\hbox{$
\;\stackrel{\textstyle<}{\sim}\;$}}

\newcommand{\bibit}[1]{\bibitem{#1}}

\newcommand{\GeV}{\,\mbox{GeV}}
\newcommand{\MeV}{\,\mbox{MeV}}
\newcommand{\matel}[3]{\langle #1|#2|#3\rangle}

\newcommand{\eod}{\end{document}}

\newcommand{\msp}[1]{\mbox{\hspace*{#1mm}~}}

\begin{document}
\begin{titlepage}
\begin{flushright}
SI-HEP-2010-07 \\UND-HEP-10-BIG\hspace*{1.5pt}04 \\[0.2cm]
\end{flushright}

\vspace{1.7cm}
\begin{center}
{\LARGE\bf 
 \boldmath $B \tto D^*$ at Zero Recoil Revisited}
\end{center}

\vspace{0.5cm}
\begin{center}
{\large {\sc Paolo Gambino}} \\[7pt]
{\sf Dipartimento di Fisica Teorica, Universit\`a di Torino \\[-2.5pt] 
\scalebox{.9}{{\rm and}} \\[-2pt] INFN Torino, I-10125 Torino, Italy} \\[30pt]
{\large{\sc Thomas Mannel, ~Nikolai Uraltsev}}\raisebox{4pt}{$^{a*}$} \\[8pt]
{\sf Theoretische Physik 1, Fachbereich Physik,
Universit\"at Siegen,  D-57068 Siegen, Germany}\\[10pt]
\scalebox{.935}{\sf  ~$^a$\,{\rm also}\, Department of Physics, 
University of Notre Dame  du Lac, Notre Dame, IN 46556 \,U.S.A.}
\end{center}

\vspace{0.8cm}
\begin{abstract}
\vspace{0.2cm}\noindent
We examine the $B \tto D^*$ form factor at zero recoil using a continuum QCD
approach rooted in the heavy quark sum rules framework.
A refined evaluation of   the radiative corrections as well as the most recent
estimates of higher order power terms together with more careful continuum
calculation are included. An upper bound on the form factor of ${\cal F} (1)
\!\lsim\! 0.93$ is derived, based on just the positivity of inelastic
contributions. A model-independent estimate of the inelastic contributions
shows they are quite significant, lowering the form factor by about $6\%$ 
or more.  This results in an unbiased estimate ${\cal F} (1)\!\approx\! 0.86$
with about three percent uncertainty in the central value.
\end{abstract}

\vfill

\noindent
\hrulefill\hspace*{290pt} \\[1pt]
$^*$\,\scalebox{.894}{On leave of absence from Petersburg Nuclear 
Physics Institute, Gatchina, St.\,Petersburg 188300, Russia}

\thispagestyle{empty}

\setcounter{page}{0}

\end{titlepage}

\newpage
\section{Introduction}
The determination of the CKM matrix element $V_{cb}$ from exclusive decays has to 
rely on calculations of the relevant form factors, which are usually defined 
as 
\begin{eqnarray}
\nonumber
\frac{d \Gamma}{d \omega} (B \tto D^* \ell \bar{\nu}_\ell) \!\!\! &=& \!\!\!
\frac{G_F^2}{48 \pi^3} |V_{cb}|^2 M_{D^*}^3 (\omega^2\!-\!1)^{1/2} P(\omega) 
 ({\cal F} (\omega))^2  \\
 \frac{d \Gamma}{d \omega} (B \tto D \: \ell \bar{\nu}_\ell) \!\!\! &=& \!\!\!
\frac{G_F^2}{48 \pi^3} |V_{cb}|^2 (M_B\!+\!M_D)^2 M_D^3  (\omega^2\!-\!1)^{3/2}  
 ({\cal G} (\omega))^2
\label{9}
\end{eqnarray}
with $\omega = v\cdot v' = E_{D^{(*)}} / M_{D^{(*)}} $ (in the $B$ rest frame),  
and where $P(\omega)$ is a known phase space factor.  Based on 
the normalization of the form factors at $\omega \!=\! 1$, $V_{cb}$ is extracted 
from an extrapolation of the data to the non-recoil point. 

In the heavy quark limit, the normalization of the form factors 
${\cal F} (1)\!=\!{\cal G} (1)\!=\! 1$ is given by heavy 
quark symmetry, and the main issue in
the $V_{cb}$ determination becomes a reliable calculation of the deviation
from the heavy quark limit. The published extractions of $V_{cb}$ along 
this route rely solely on the lattice calculations currently cited as \cite{lattice}
\begin{eqnarray}
\nonumber
&& \mathcal{F}(1) = 0.921\pm 0.024 \\ 
&& \mathcal{G}(1) = 1.074 \pm 0.018 \pm 0.016
\label{48}
\end{eqnarray} 
However, based on the dynamic heavy quark expansion in
Minkowski space, it has been argued that 
larger deviations from the symmetry limit for $\mathcal{F}(1)$ 
are natural in continuum   QCD \cite{ioffe}; these have been further
supported by the arguments \cite{chrom} exploiting the relatively
small kinetic expectation value $\mu_\pi^2$ extracted from the fits to
inclusive $B$ decays. The same line of reasoning led to a rather
precise estimate of $\mathcal{G}(1)$ in $B\tto D$ \cite{BPS}, showing
significantly {\sl smaller} deviations from unity compared to
Eq.~(\ref{48}).

Recent years were of primary importance for 
heavy flavor physics. Along with advances in 
theory, many nontrivial nonperturbative predictions were
verified with high precision, and heavy quark parameters
experimentally extracted in accord with prior theoretical expectations;
certain predictions were indirectly confirmed in dedicated lattice 
calculations. All this raised the credibility of OPE-based methods and 
favored an early onset of the short-distance expansion instrumental for
high-precision predictions; confidence rose in the assumptions
underlying dynamic treatment of the nonperturbative physics in heavy quarks. 
This progress warrants a critical re-examination of the form factors. 
The goal is to incorporate the accumulated knowledge and to shift the focus from
merely establishing the scale of the deviations from the heavy quark symmetry
limit towards obtaining a refined estimate with a motivated error  
assessment.

In the present note we discuss the $B \tto D^*$ transition at zero recoil using
a dynamic QCD approach inspired by  
the original treatment of the zero-recoil sum rules for heavy
flavor transitions. The details of the analysis will be presented in
the extended publication \cite{f0long}.

\section{Zero Recoil Sum Rule in QCD} 
\begin{figure}[t]
\begin{center}
\mbox{\epsfig{file=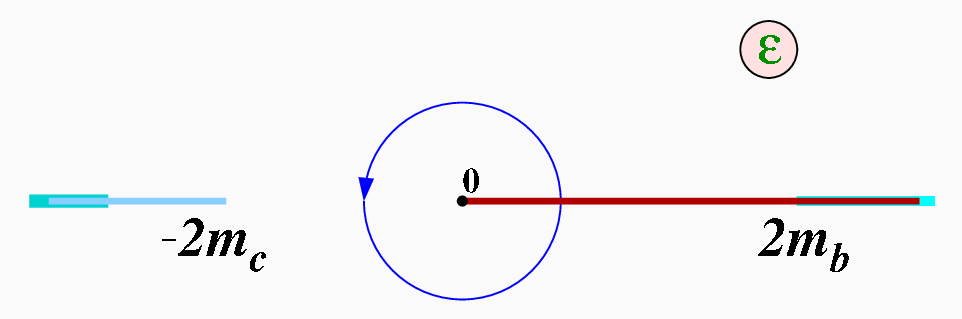}}\vspace*{-15pt}
\end{center}
\caption{The analytic structure of $T^{\rm zr}(\varepsilon)$ 
and the integration contour yielding the sum rule. Distant cuts are
shown along with the physical cut. The radius of the circle is 
$\varepsilon_{M}$.}
\label{fig1}
\end{figure}

We consider the zero-recoil ($\vec q \!=\!0$) forward scattering amplitude
$T^{\rm zr}(\varepsilon)$ of the flavor-changing axial current $\bar{c}\vec\gamma
\gamma_5 b$ off a $B$ meson at rest:
\beq
T^{\rm zr}(\varepsilon)=\int\! {\rm d}^3x
\int\! {\rm d}x_0\; e^{-i x_0 (M_B - M_{D^*} - \varepsilon) }
\frac{1}{2M_B} \matel{B}{\mbox{$\frac{1}{3}$}\:{ iT}\,
  \bar{c}\gamma_k\!\gamma_5b(x)\:\bar{b}\gamma_k\!\gamma_5 c(0)}{B}\, , 
\label{80}
\eeq
where $\varepsilon$ is the excitation energy above $M_{D^*}$
in the $B \tto X_c $ 
transition (the point $\varepsilon=0$ corresponds to 
the elastic $B\to D^*$ transition). 
The amplitude $T^{\rm zr}(\varepsilon)$ is
an analytic function of $\varepsilon$ and has a physical decay cut at
$\varepsilon\!\ge\!0$, and other distant singularities. The 
analytic structure of $T^{\rm zr}(\varepsilon)$ 
is shown in Fig.~\ref{fig1}. 

The contour integral
\beq
I_0(\varepsilon_{M})= -\frac{1}{2\pi i}\;
\oint_{\raisebox{-3pt}{\hspace*{-5pt}\scalebox{.7}{$|\varepsilon|\!=\!
\varepsilon_M$}}\hspace*{-5pt} }T^{\rm zr}(\varepsilon) \,{\rm d}\varepsilon
\label{82}
\eeq
with the contour running counterclockwise from the upper side of the
cut, see Fig.~\ref{fig1}, leads to the sum rule involving ${\mathcal
F}^2(1)$. Using the analytic properties of $T^{\rm zr}(\varepsilon)$
the integration contour can be shrunk onto the decay cut; the
discontinuity there is related to the weak transition amplitude
squared of the axial current into the final charm state with mass
$M_X\!=\!M_{D^*}\!+\!\varepsilon$. Separating out explicitly the
elastic transition contribution $B\tto D^*$ at $\varepsilon\!=\!0$ we
have
\beq
I_0(\varepsilon_{M})= {\cal F}^2(1)+w_{\rm inel}(\varepsilon_{M}), 
\qquad w_{\rm inel}(\varepsilon_{M}) \!\equiv\!
\frac{1}{2\pi i}\;  \int_{\varepsilon>0}^{\varepsilon_{M}}
{\rm disc}\,  T^{\rm zr}(\varepsilon) \,{\rm d}\varepsilon\,,
\label{84}
\eeq
where $w_{\rm inel}(\varepsilon_{M})$ is related to the sum of the
differential decay probabilities into the excited states with mass up to
$M_{D^*}\!+\!\varepsilon_{M}$ in the zero recoil kinematics. 

The OPE allows us to calculate the amplitude in (\ref{80}) -- and hence
$I_0(\varepsilon_{M})$ -- in the short-distance expansion provided
$|\varepsilon|$ is sufficiently large compared to   the ordinary 
hadronic mass scale. It
should be noted that strong interaction corrections are driven not only by
$|\varepsilon|$, but also by the proximity to distant singularities.
Therefore, $\varepsilon_{M} $ cannot be taken too large either, and the hierarchy
$\varepsilon_{M}\!\ll\!2m_c$ has to be observed.

The sum rule Eq.~(\ref{84}) can be cast in the form
\beq
 {\cal F}(1)= \sqrt{I_0(\varepsilon_{M})\!-\!w_{\rm inel}(\varepsilon_{M})}
\label{86}
\eeq
which is the master identity for the considerations to follow. Since 
$w_{\rm  inel}(\varepsilon_{M})$    is strictly positive, we get an upper bound 
on the form factor 
\beq
{\cal  F}(1) \le \sqrt{I_0(\varepsilon_{M})}
\label{86a}
\eeq
which relies only on the OPE calculation of $I_0$. Note that this bound
depends on the parameter $\varepsilon_{M}$, while (\ref{86}) is independent of
$\varepsilon_{M}$ since the dependence in $I_0$ and 
$w_{\rm inel}$   cancel.  Furthermore, including an estimate
of $w_{\rm inel}(\varepsilon_{M})$ we obtain an evaluation of 
${\cal F}(1)$.

The correlator in (\ref{80}) can be computed using the OPE, resulting in
an expansion of $T^{\rm zr}(\varepsilon)$ in inverse powers of the masses $m_c$
and $m_b$. This results in the
corresponding expansion of $I_0 (\varepsilon_{M})$. This OPE takes the
following general form
\bea \label{90}
I_0(\varepsilon_{M}) &=& \xi_A^{\rm pert}(\varepsilon_{M},\mu) + 
\sum_{k} C_k(\varepsilon_{M},\mu)
\, \frac{\mbox{$\frac{1}{2M_B}$}\matel{B}{O_k}{B}_\mu}{m_Q^{d_k-3}} \\ 
\nonumber 
&=&
\xi_A^{\rm pert}(\varepsilon_{M},\mu)-\Delta_{1/m_Q^2}(\varepsilon_{M},\mu)
-\Delta_{1/m_Q^3}(\varepsilon_{M},\mu)-
\Delta_{1/m_Q^4}(\varepsilon_{M},\mu)-...
\eea
where $O_k$ are local $b$-quark operators $\bar{b}...b$ of increasing dimension
$d_k\!\ge\!5$,  $C_k(\mu)$ are Wilson coefficients for power-suppressed terms, 
and $\xi_A^{\rm pert}$ is the short-distance  renormalization (corresponding to the 
Wilson coefficient of the unit operator), which is unity at tree level. 
We have also introduced a Wilsonian cutoff $\mu$ used to separate long and short 
distances. The complete result, of course, does not depend 
on $\mu$ since the $\mu$-dependence cancels between the Wilson coefficients 
and the matrix elements of the operators. 
At tree level $\Delta$ does not depend on $ \varepsilon_{M}$. 
The choice of $\mu$ is subject to the same general constraints as that 
of $\varepsilon_M$, and 
it is therefore convenient to choose $\mu\!=\!\varepsilon_{M}$. 

\subsection{Perturbative corrections}

The perturbative renormalization $\xi_A^{\rm  pert}(\mu)$
can be expanded in power series in
$\alpha_s$. We use the Wilsonian OPE and benefit from well-behaved  
perturbative series for $\xi_A^{\rm  pert}(\mu)$.
The exact form of the perturbative coefficients 
depends on the definition chosen for higher-dimension operators;
we adopt the often used {\it kinetic} scheme \cite{fivedipole}. 

In one-loop perturbative calculations there is a simple connection
between the normalization point of the heavy quark operators in the
kinetic scheme and the hard cut-off on the gluon momentum in the
diagram. This allows to obtain the analytic expression for
$\xi_A^{\rm pert}(\mu)$ to this order even without explicit
calculation of the Wilson coefficients $C_k$ in Eq.~(\ref{90}). The
expression is rather lengthy and will be presented in  
Ref.~\cite{f0long}. By the same trick one also obtains all
higher-order BLM corrections by performing the 
one-loop calculations with massive gluon \cite{imprec}.

A similar argument does not apply to non-BLM corrections starting
$\alpha_s^2$ where $\varepsilon_M$-dependence of $\xi_A^{\rm pert}$ 
has to be determined expanding in $1/m_Q$; for ${\cal O}(\alpha_s^2)$
corrections this was done in Ref.~(\cite{xi2}) through order
$1/m_Q^2$. The corresponding coefficient was found to be
small numerically, which suggests that omitted terms  $\propto \alpha_s^2
\varepsilon_M^3/m_Q^3$ and higher should not produce a significant change.

Perturbative corrections to $\xi_A^{\rm pert}(\varepsilon_M )$ appear
to be small for practical values of $\varepsilon_M $ between $0.6\GeV$
and $1\GeV$. Taking, for instance, $\varepsilon_M\!=\!0.75\GeV$,
$m_c\!=\!1.2\GeV$, $m_b\!=\!4.6\GeV$, $\alpha_s(m_b)\!=\!0.22$ we get 
the numeric estimates at different orders 
\beq
\sqrt{\xi_A^{\rm pert}}=1-0.019+(0.007-0.004)+0.0045 +...
\label{96}
\eeq
Here the first term is the tree value, second is ${\cal O}(\alpha_s)$
evaluated with $\alpha_s\!=\!0.3$, the next pair of values show the shift
upon passing to the  ${\cal O}(\alpha_s^2)$ order (positive for the BLM part
and negative from the non-BLM contribution); the last term shows
$\beta_0^2\alpha_s^3$ term as an estimate of even higher-order perturbative
corrections.
\begin{figure}[h]
 \begin{center}
\mbox{\epsfig{file=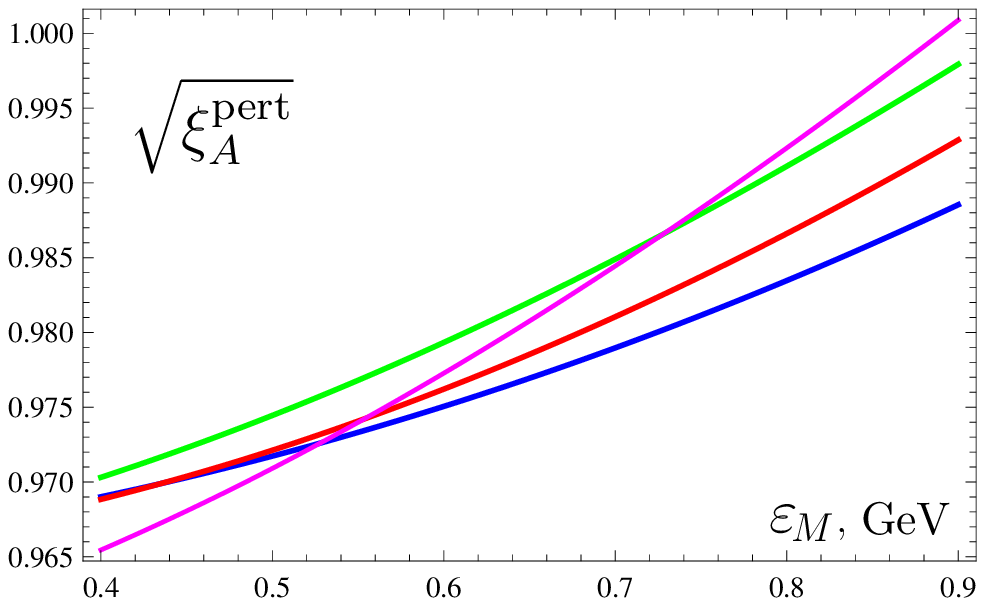,width=6.5cm}} \hfill 
\mbox{\epsfig{file=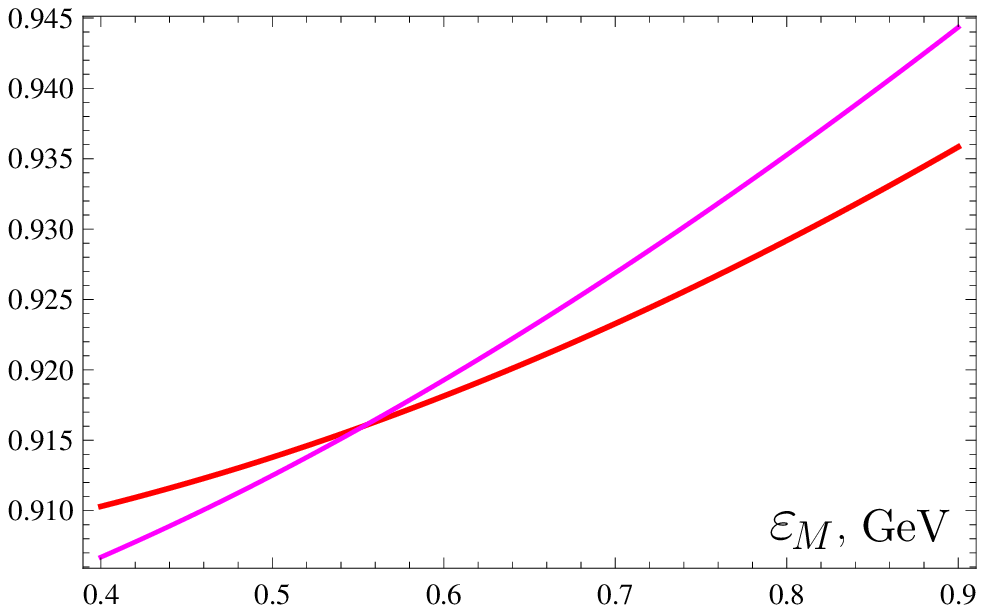,width=6.5cm}} 
\vspace*{-15pt}
 \end{center}
\caption{ \small {\bf Left:}\,\,\,\, $\sqrt{\xi_A^{\rm pert}}$ to order 
$\alpha_s$ (blue),
    including $\beta_0\alpha_s^2$ (green), full ${\cal O}(\alpha_s^2)$ (red)
    and including  \newline
\hspace*{91pt}$\beta_0^2\alpha_s^3$ (magenta), assuming 
$\alpha_s^{\overline{{\rm MS}}}(m_b)\!=\!0.22$, $m_c\!=\!1.2\GeV$ and
$m_b\!=\!4.6\GeV$. \newline
\hspace*{50pt}{\bf Right:}\,  Upper 
bound (\ref{86a}) on $\mathcal{F}(1)$ depending on
$\varepsilon_M$, with or without $\beta_0^2\alpha_s^3$
term. A \hspace*{91pt}fixed $ \Delta\!=\!0.11$ is used; assuming 
the perturbative evolution of $\Delta$ with $\varepsilon_M$ 
would \hspace*{91pt}flatten the dependence.}
\label{xipert}
\end{figure}
Fig.~\ref{xipert} shows the dependence on $\varepsilon_M$ of these predictions 
for  $\sqrt{\xi_A^{\rm pert}}$. 
In particular, taking the full two-loop result as the central estimate we find
\begin{equation}
\sqrt{\xi_A^{\rm pert}(\mbox{0.75\GeV})} = 0.985 \pm 0.01\,;
\end{equation} 
we will use  $\varepsilon_M\!=\!0.75$GeV in the 
subsequent discussion.  We emphasize that the numeric stability
applies only to the perturbative renormalization factor in the
Wilsonian OPE, and the quoted values refer to the specific
renormalization scheme (kinetic) adopted in the analysis.

\subsection{Power corrections}

The leading power corrections to $I_0$ were calculated in
Refs.~\cite{vcb,optical} to order $1/m_Q^2$ and to order $1/m_Q^3$ in
Ref.~\cite{rev} and read
\bea
\nonumber
\Delta_{1/m^2} &\msp{-5}=\msp{-5}& \frac{\mu_G^2}{3m_c^2} +
\frac{\mu_\pi^2 \!-\!\mu_G^2 }{4}
\left(\frac{1}{m_c^2}+\frac{2}{3m_cm_b}+\frac{1}{m_b^2}
\right), \\
\Delta_{1/m^3} &\msp{-5}=\msp{-5}&
\frac{\rho_D^3 - \frac{1}{3}
\rho_{LS}^3}{4m_c^3}\;+\;
\frac{1}{12m_b}\left(\frac{1}{m_c^2}+\frac{1}{m_c m_b} +\frac{3}{m_b^2}\right)
\,(\rho_D^3 +
\rho_{LS}^3)\,.
\label{103}
\eea
The nonperturbative parameters  $\mu_\pi^2$, $\mu_G^2$, $ \rho_D^3$ and
$\rho_{LS}^3$ all depend on the hard Wilsonian cutoff.  
The $\varepsilon_M$-dependence of $\xi_A^{\rm pert}$ is linked to the power-like 
scale dependence of the nonperturbative matrix elements through their 
mixing with lower-dimension operators. 

In the numerics we use the values
$\mu_\pi^2 (0.75\GeV)\!=\!0.4\GeV^2$,  $\rho_D^3
(0.75\GeV)\!=\!0.15\GeV^3$,
while for the quark masses $m_c\!=\!1.2\GeV$ and $m_b\!=\!4.6\GeV$ (the
scale dependence of the latter plays a role here only at the level
formally beyond the accuracy of the calculation).
The dependence on $\mu_G^2$ and on  $\rho_{LS}^3$ is minimal and their
precise values do not matter; we use  for them  $0.3\GeV^2$ and
$-0.12\GeV^3$, respectively.  We then get
\beq
\Delta_{1/m^2}= 0.091, \qquad \Delta_{1/m^3}= 0.028\,;
\label{106}
\eeq
If we employ the values of the OPE parameters
extracted from a fit to inclusive semileptonic and radiative decay
distributions \cite{HFAG,fit}, we find a consistent result
\beq
 \Delta_{1/m^2}+\Delta_{1/m^3} = 0.11 \pm 0.03 \, .
\label{110a}
\eeq

An important question is how well the power expansion for the sum
rule converges. Recently, the OPE for the semileptonic
$B$-meson structure functions has been extended to order $1/m_Q^4$ and
$1/m_Q^5$ \cite{icsieg,hiord}. Combined with the estimates \cite{ic} of
the corresponding expectation values discussed in Ref.~\cite{hiord}, this
leads to
\beq
\Delta_{1/m^4}\simeq -0.023, \qquad \Delta_{1/m_Q^5}\!
\simeq\!-0.013.
\label{108}
\eeq
We then observe that the
power series for $I_0$ appears well-behaved at the required level of
precision.
For what concerns the  loop corrections to $\Delta$,
the ${\cal O}(\alpha_s)$ correction to the Wilson coefficient for
the kinetic operator in Eq.~(\ref{103}) was calculated in
Ref.~\cite{xi2} and turned out numerically insignificant. At 
${\cal O}(1/m_Q^3)$, even if radiative corrections change the coefficient
for the Darwin term by $30\%$ the effect on the sum rule would still be
small.

Taking into account all the available information, our estimate for
 the total power correction at $\varepsilon_M\!=\!0.75\GeV$ is 
\beq
\Delta = 0.105
\label{110}
\eeq
with a $0.015$ uncertainty due to higher orders. On theoretical grounds,
larger values of  $\mu_\pi^2$ and/or $\rho_D^3$ are actually favored; they tend
to increase  $\Delta$.
Combining the above with the perturbative corrections we arrive at an
estimate for $I_0$ and, according
to Eq.~(\ref{86a}) at a bound for the form factor, which in terms of central values
at $\varepsilon_M \!=\!0.75\GeV$ is
\beq
{\cal F}(1) < 0.93 \,.
\label{139}
\eeq

As stated above, the upper bound in Eq.~(\ref{86a}) depends on $\varepsilon_M$,
see Fig.~\ref{xipert}, becoming stronger for smaller $\varepsilon_M$. 
It is  advantageous to choose  the minimal value of
$\varepsilon_M$ for which the OPE-based short-distance  expansion 
of the integral (\ref{82}) for $I_0(\varepsilon_M)$ sets in. This directly
depends on how low one can push the renormalization scale $\mu$ 
while still observing the expectation values actual $\mu$-dependence
in the kinetic
scheme approximated by the perturbative one. Since in this scheme
$\mu_\pi^2(\mu)\!\ge \!\mu_G^2(\mu)$ holds for arbitrary $\mu$, in essence this
boils down to the question at which scale 
$\mu_{\rm min}$ the spin sum rule and the one
for $\mu_G^2$ get approximately saturated, e.g.\  
$\mu_G^2(\mu_{\rm min})\!\simeq\! 0.3\GeV^2$. 
The only vital assumption in the analysis is that the
onset of the short-distance regime is not unexpectedly delayed in actual QCD 
and hence does not require $\varepsilon_M \!>\! 1\GeV$. This principal question 
can and should be verified on the lattice. This will complement already
available evidence from 
preliminary lattice 
data \cite{tau32} as well as from the  successful experimental 
confirmation \cite{Belle} in nonleptonic $B$ decays of the 
predicted $3/2$-dominance.

\subsection{Estimate of the inelastic contribution 
\boldmath $w_{\rm inel}$ }

On general grounds $w_{\rm inel}$ is expected to be comparable to  
the power correction $\Delta$ considered above. 
To actually estimate it we consider another contour integral
\beq
I_1(\varepsilon_M)= -\frac{1}{2\pi i}\;
\oint_{\raisebox{-4pt}{\hspace*{-4pt}\scalebox{.7}{$|\varepsilon|\!=\!\varepsilon_M$}}\hspace*{-5pt} }
T^{\rm zr}(\varepsilon) \, \varepsilon\,{\rm d}\varepsilon 
\label{120}
\eeq
for which we can write 
\beq
 w_{\rm inel}(\varepsilon_M) = \frac{I_1(\varepsilon_M)}{\tilde\varepsilon}
\label{122}
\eeq
where $\tilde\varepsilon$ is the  average excitation energy (it depends on
$\varepsilon_M$). The integral is expected to be dominated by the lowest radial
excitations of the ground state, with $\tilde\varepsilon \!\approx\!
\epsilon_{\rm rad}\!\approx\!700\MeV$. 

$I_1(\varepsilon_M)$ can also be calculated 
in the OPE  \cite{optical}; the result including 
$1/m_Q^2$ terms reads 
\beq
I_1= \frac{-(\rho_{\pi G}^3+\rho_A^3)}{3m_c^2}
+ \frac{-2\rho_{\pi\pi}^3-\rho_{\pi G}^3}{3m_c m_b}+
\frac{\rho_{\pi\pi}^3+\rho_{\pi G}^3+\rho_S^3+\rho_A^3}{4} \left(
\frac{1}{m_c^2}+  \frac{2}{3m_c m_b}+ \frac{1}{m_b^2} \right)
\label{124}
\eeq
where the non-local zero momentum transfer correlators $\rho_{\pi
\pi}^3$, $\rho_{\pi G}^3$, $\rho_S^3$ and $\rho_A^3$ are defined in
\cite{optical}.  They can be estimated along the lines described in
\cite{hiord}, based on saturation by the appropriate intermediate
states. We shall defer the details of this estimate to \cite{f0long}.
Here we note that only the
first term survives in the BPS limit \cite{BPS}, where it is positive;
the second and third terms are of first and second order in the
deviation from the BPS limit, respectively. The last term is positive
being the correlator of  two identical operators
$\bar{b}(\vec\sigma \vec\pi)^2 b$. Since the middle term comes with a
small coefficient $1/3m_c m_b$, the expression has only a shallow
minimum where the first term is decreased by less than $10\%$; the sum
of the last two terms becomes larger than that only if it is
positive. On the other hand, the combination  
\beq
-(\rho_{\pi G}^3+\rho_A^3) + \rho_{LS}^3
\label{160}
\eeq
determines the hyperfine splitting to order $1/m_Q^2$ and can be constrained
from the observed masses of $B^{(*)}$ and  $D^{(*)}$ mesons.
We finally obtain
\beq
I_1 (\varepsilon_M) \gsim \frac{0.48\GeV^3}{3m_c^2}
\label{174}
\eeq
with some uncertainty from perturbative corrections, 
implying for $\tilde\varepsilon \!=\! \epsilon_{\rm rad}\!\simeq\!
700\MeV$
\beq
w_{\rm inel} \approx \frac{I_1}{\epsilon_{\rm rad}}\gsim 0.13\;.
\label{176}
\eeq

This estimate is derived at leading order in $1/m_Q$ and may be
corrected by higher-order terms by as much as $30\%$. 
We observe that $w_{\rm inel}$ 
is similar in size 
to the power term in $I_0$ and even exceeds it. Using (\ref{176}) 
at face value we arrive at our estimate for the expected value of
the form factor 
\beq 
{\cal F}(1) \lsim 0.86.
\eeq 

\begin{figure}[t]
\vspace*{2pt}
 \begin{center}
\hfill \mbox{\epsfig{file=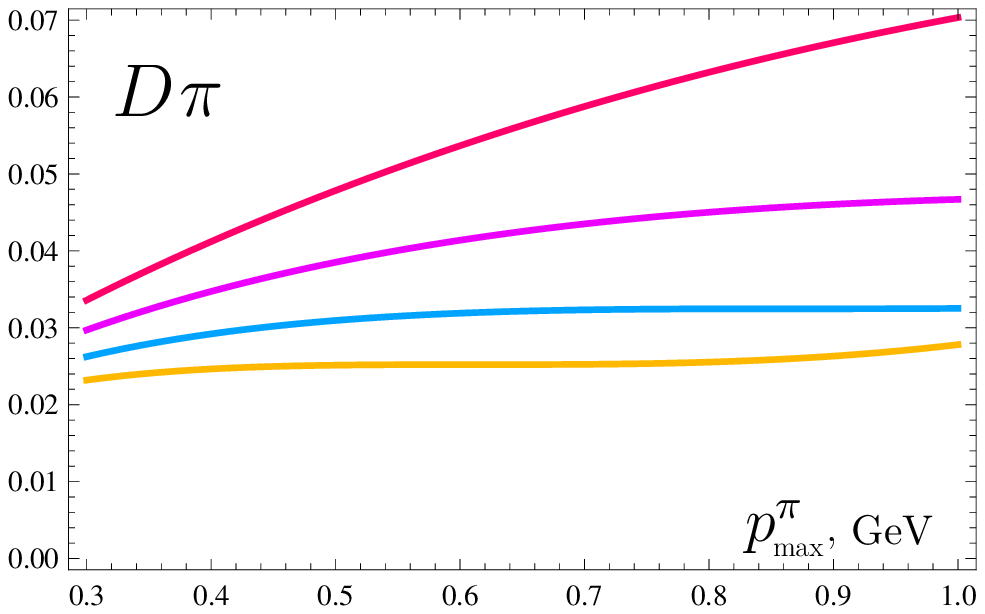,width=5.5cm}} \hfill\hfill
\raisebox{-1pt}{\mbox{\epsfig{file=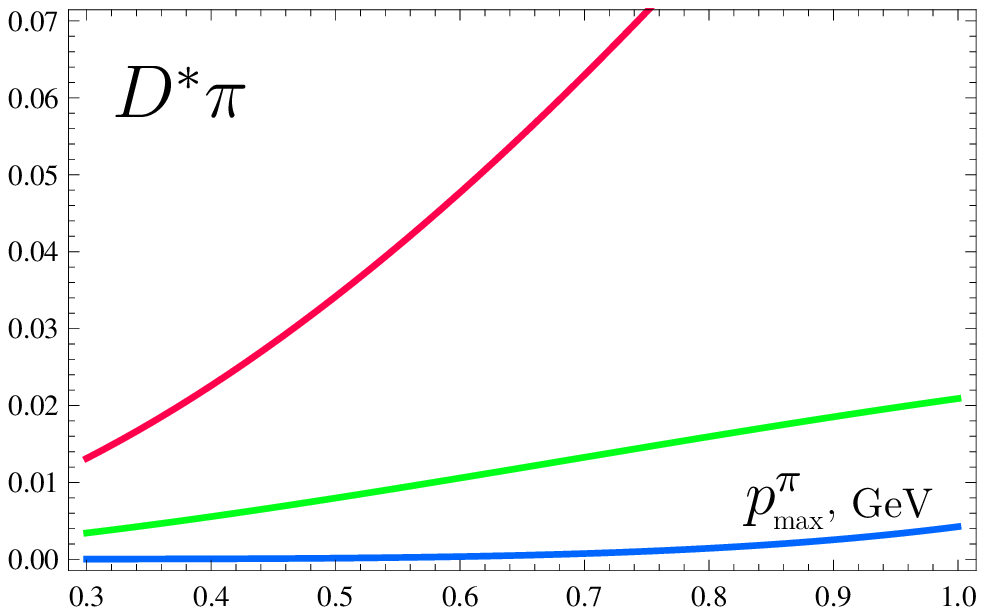,width=5.5cm}}}\hfill ~
\vspace*{-12pt}
 \end{center}
\caption{ \small
Non-resonant $D\pi$  and $D^*\pi$  contribution to $w_{\rm inel}$ 
at $g_{D^*D\pi}\!=\!4.9\GeV^{-1}$ corresponding to 
$\Gamma_{D^{*+}}\!=\!96\,{\rm keV}$. The plots show, from bottom to top, 
$\frac{g_{B^*B\pi}}{g_{D^*D\pi}}\!=\!1$, $0.8$,
 $0.6$ and $0.4$, and $\frac{g_{B^*B\pi}}{g_{D^*D^*\pi}}\!=\!1$, $0.8$ and
 $0.6$, respectively
 }
\label{contin}
\end{figure}

The quasi-resonant states are expected to dominate $w_{\rm inel}(\mu)$
at intermediate $\mu \!\approx\! 1\GeV$; the continuum contribution to
$w_{\rm inel}(\mu)$ is parametrically $1/N_c$-suppressed and usually
smaller.  The $D^{(*)}\pi$ continuum 
can independently be evaluated in the soft-pion
approximation. It turns out that numerically the dominant effect
originates from the heavy quark symmetry breaking difference between
the $B^*B\pi$, $D^*D\pi$ and $D^*D^*\pi$-couplings which until
recently has not been accounted for in this context,\footnote{The
$D^*\!D^*\!\pi$ channel has not been previously considered while
generally required by the heavy quark symmetry.} although is expected
to be significant \cite{khod}. The result is shown in
Fig.~\ref{contin}; it depends on the upper cutoff in the pion momentum
$p^\pi_{\rm max}$ marking the end of the soft continuum domain for
$D^{(*)}\pi$, presumably somewhat below $\epsilon_{\rm rad}$. We
expect about $5\%$ combined yield for $\Gamma_{D^{*+}}\!=\!96\,{\rm
keV}$, i.e.\ about a third of the overall $w_{\rm inel}$ in
Eq.~(\ref{176}) in accord with the $1/N_c$ arguments. This
contribution alone would lower the upper bound in Eq.~(\ref{139}) by
$0.025$.

\thispagestyle{plain}

\section{Conclusions}

The direct OPE-based $1/m_Q$ expansion of the zero recoil form factor which we
have analyzed through the sum rule for the correlator of the
zero-recoil axial currents yields the unbiased estimate 
${\cal F}(1)\!\approx\!0.86$, and suggests a lower bound 
${\cal F}(1)\! <\!0.93$. 
All the values refer to pure QCD form factors and
exclude possible electromagnetic effects, in particular the universal
short-distance semileptonic enhancement factor of $1.007$.

Since a charm mass expansion is involved, the precision of our estimates 
is limited. Accuracy-wise two aspects should be kept distinct.
On the one hand,  the 
heavy quark parameters enter our analysis, and in particular
$\mu_\pi^2$, $\rho_D^3$ and $m_c$. Their determination  from inclusive measurements
will soon be improved  thanks to refined theoretical calculations.
On the other hand,  even if these QCD parameters were accurately known,
sharpening the upper bound and  the estimate of ${\cal F}(1)$ would
require additional nonperturbative input. 
To some extent the uncertainties may be reduced by
dedicated measurements in semileptonic decays or by direct lattice
calculation of the relevant heavy quark parameters. In our opinion, the latter can
be done in a straightforward way, which will be discussed elsewhere.
 
In our numerical estimates we used the
lowest values of $\mu_\pi^2$ and $\rho_D^3$ 
consistent with the theoretical bounds; larger values
would typically further lower both the upper
bound and the central expectation for ${\cal F}(1)$. 
We estimate the perturbative and non-perturbative uncertainties in our 
approach  to be about $\pm 0.01_{\rm pt}$ and 
$\pm 0.02_{\rm  np}$ ($\pm 0.01_{\rm np}$ for the upper bound), 
mostly related to the
unavoidable truncation of the $1/m_c$ expansions. The uncertainty in the
prediction is dominated by $w_{\rm inel}$, is not symmetric, and allows
for larger deviations towards {\sl lower} values of ${\cal F}(1)$. 

There is an alternative way of determining $|V_{cb}|$, using inclusive
semileptonic decays. It rests on the heavy quark expansion in $1/m_b$
and on the OPE which, in this case, should be considered a mature
tool. Recent estimates of higher orders in $1/m_Q$ as well as in
$\alpha_s$ did not show signs of failure of the heavy quark expansion
in this application. Hence one can employ the value of $V_{cb}=(41.54
\pm 0.44 \pm 0.58 ) \times 10^{-3}$ from the inclusive decays
\cite{HFAG} and determine the form factors from the data on $B\to
D^{(*)}\ell \nu$ decays; this yields
\begin{eqnarray}
\label{204}
&& \mathcal{F}(1) = 0.86 \pm 0.02 \\
\label{205}
&& \mathcal{G}(1) = 1.02 \pm 0.04
\end{eqnarray}  
The value for $\mathcal{G}(1)$ is in a perfect agreement with the HQE
prediction \cite{BPS}; $\mathcal{F}(1)$ hits the central value of our estimate.
Both form factors in Eq.~(\ref{204}) are noticeably below the quoted
lattice values Eqs.~(\ref{48}); this explains 
the observed tension between the inclusive and exclusive
determinations of $|V_{cb}|$.

Our analysis suggests that the lattice determinations of both form
factors are systematically high. The central values of the currently
quoted lattice ${\cal F}(1)$ is very close to our upper bound. Lattice
calculations -- provided they accommodate the experimentally measured
$B$-meson expectation values -- seem to imply an   extremely small
inelastic contribution, which is a priori unnatural. Moreover,
it appears in contradiction with the large non-local correlators
encountered to order $1/m_Q^2$. The estimate of the non-resonant
soft-pion $D^{(*)}\pi$ rates also confirms this assessment.  We
conclude that values for ${\cal F}(1)$ in excess of $0.9$ would be
consistent with unitarity and the short-distance expansion of QCD only
with rather contrived assumptions. Values of ${\cal F}(1)$ larger than
$0.93$ should be viewed as violation of unitarity assuming that usual
short-distance expansion in QCD works.

With the unbiased estimate based on the lower-end $\mu_\pi^2$ and $\rho_D^3$
yielding ${\cal F}(1)\!\approx\! 0.86$ -- or even smaller for larger
$\mu_\pi^2, \,\rho_D^3$ -- we find a surprisingly good agreement (probably,
somewhat accidental) between the exclusive and inclusive approaches. In view
of the inherent $1/m_c$ expansion for ${\cal F}(1)$ and of the 
approximations used for proliferating hadronic expectation values, 
  matching the theoretical precision already attained for $V_{cb}$ from the
inclusive fits does not look probable. Nevertheless, additional experimental
and/or lattice input would make the suggested  $3\%$ uncertainty
interval more robust.

We estimate that the contribution of excited radial states with mass 
below $3\GeV$ constitutes about $10\%$ or more of the total yield. This refers only 
to the zero-recoil kinematics and only to the axial current; it does not
include $P$-wave states. This fraction may even be larger
when applied to the full phase space. This suggests that the observed `broad state'
yield in this mass range routinely attributed to the $\frac{1}{2}$ $P$-wave
excitations is actually dominated by  states with different quantum
numbers, thus resolving the `$\frac{1}{2}\!>\!\frac{3}{2}$ puzzle'. 
The suppression of the broad $P$-wave yield was predicted based
on the spin sum rules and confirmed indirectly in nonleptonic $B$
decays \cite{Belle}; a recent discussion can be found in Ref.~\cite{bigi1232}.

\subsection*{Acknowledgments}

We gratefully acknowledge discussions with Ikaros Bigi and invaluable
help from Sascha Turczyk regarding higher-order power corrections. We
thank the Galileo Galilei Institute for Theoretical Physics for the
hospitality and the INFN for partial support during the completion of
this work. The study enjoyed a partial support from the NSF grant
PHY-0807959 and from the RSGSS-65751.2010.2 grant.   PG is supported in
part by a EU's Marie-Curie Research Training Network 
under contract  MRTN-CT-2006-035505 (HEPTOOLS). TM is
partially supported by the German research foundation DFG under
contract MA1187/10-1 and by the German Ministry of Research (BMBF),
contracts 05H09PSF.

\end{document}